\documentclass{article}
\usepackage{arxiv}

\usepackage[utf8]{inputenc} 
\usepackage[T1]{fontenc}    
\usepackage{hyperref}       
\usepackage{url}            
\usepackage{booktabs}       
\usepackage{amsfonts}       
\usepackage{nicefrac}       
\usepackage{microtype}      
\usepackage{lipsum}
\usepackage{float}
\usepackage{graphicx}
\usepackage{multirow}
\usepackage{array}
\usepackage{comment}
\usepackage{longtable}
\usepackage{ragged2e}
\usepackage{subfig}
\usepackage[table]{xcolor}  
\usepackage{geometry}
\usepackage{multicol}
\usepackage{amsmath}
\usepackage{pifont}
\title{Metrology and Manufacturing-Integrated Digital Twin (MM-DT) for Advanced Manufacturing: Insights from CMM and FARO Arm Measurements}

\author{
  Hamidreza Samadi \\
  Industrial and Systems Engineering\\
  University of Oklahoma\\
  Norman, Oklahoma-73069 \\
  \texttt{hamidreza.samadi@ou.edu} \\
   \And
  Md Manjurul Ahsan \\
  Industrial and Systems Engineering\\
  University of Oklahoma\\
  Norman, Oklahoma-73071\\
  \texttt{ahsan@ou.edu} 
   \And 
  Shivakumar Raman \\
  Industrial and Systems Engineering\\
  University of Oklahoma\\
  Norman, Oklahoma-73071\\
  \texttt{raman@ou.edu} 
   \And
}

\begin{document}
\maketitle

\begin{abstract}
Metrology, the science of measurement, plays a key role in Advanced Manufacturing (AM) to ensure quality control, process optimization, and predictive maintenance. However, it has often been overlooked in AM domains due to the current focus on automation and the complexity of integrated precise measurement systems. Over the years, Digital Twin (DT) technology in AM has gained much attention due to its potential to address these challenges through physical data integration and real-time monitoring, though its use in metrology remains limited. Taking this into account, this study proposes a novel framework, the Metrology and Manufacturing-Integrated Digital Twin (MM-DT), which focuses on data from two metrology tools, collected from Coordinate Measuring Machines (CMM) and FARO Arm devices. Throughout this process, we measured 20 manufacturing parts, with each part assessed twice under different temperature conditions. Using Ensemble Machine Learning methods, our proposed approach predicts measurement deviations accurately, achieving an R\textsuperscript{2} score of 0.91 and reducing the Root Mean Square Error (RMSE) to 1.59 µm. Our MM-DT framework demonstrates its efficiency by improving metrology processes and offers valuable insights for researchers and practitioners who aim to increase manufacturing precision and quality.

\end{abstract}


\keywords{Digital Twins \and Metrology \and Coordinate Measuring Machine (CMM)\and FARO Arm \and Industry 4.0 \and Quality Control \and Predictive Maintenance \and Machine Learning \and Advanced Manufacturing}

\section{Introduction}\label{sec1}
A Digital Twin (DT) serves as a digital copy of physical assets, allowing for simulations, data analysis, and immediate adjustments. In Advanced Manufacturing (AM), DTs help manufacturers improve efficiency, cut down on delays, and improve the quality of products~\cite{ahsan2024digital, ouahabi2024leveraging}. By using DTs, manufacturers can quickly prototype complex designs, make iterative updates, and adjust processes on the fly~\cite{egon2024digital, alfred2024implementation, Hamel2024PMIDTLD}. These capabilities enable companies to design, test, and refine products with greater precision~\cite{filippova2024use, arumugam2024analysis}.

Metrology, which deals with accurate measurement, plays a crucial role in ensuring products meet their design standards. This discipline focuses on measuring dimensions precisely, which ensures consistency across the production process. In manufacturing, precise measurements help companies boost efficiency, improve product performance, and maintain high reliability~\cite{kimothi2001uncertainty}.

\begin{figure}[h!]
    \centering
    \includegraphics[width=0.5\linewidth]{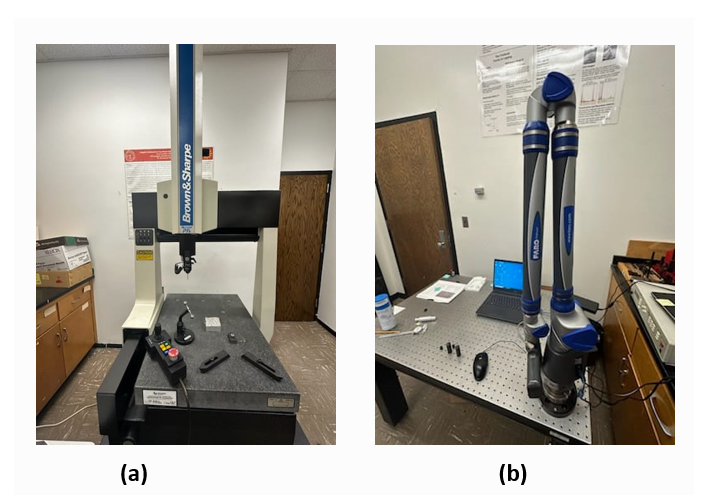}
    \caption{Illustration of (a) CMM and (b) FARO Arm machine used during this study for precise dimensional measurement and real-time inspection of parts, respectively.}
    \label{fig:cmmfaro}
\end{figure}

Manufacturers often use Coordinate Measuring Machines (CMM) and FARO Arms to check geometric properties and ensure parts meet precise dimensions~\cite{olu-lawal2024role}. CMMs excel at verifying close tolerances and providing in-line measurements, which helps maintain real-time control~\cite{pant2023role}. On the other hand, FARO Arms offer the flexibility needed to measure larger components and complex shapes, especially in situations where traditional CMMs fall short~\cite{pant2023role}. Their portability allows engineers to take measurements directly at the production site, avoiding the need to move parts and speeding up the process. This versatility not only cuts down on defective parts but also improves efficiency, resulting in cost savings. However, challenges like calibration issues and measurement uncertainties remain. These challenges are addressed through predictive maintenance and autonomous quality control, where technological integration such as Internet of Things (IoT), Artificial Intelligence (AI), and metrology data from physical assets can be monitored and analyzed using DTs~\cite{singh2022integration}. Figure~\ref{fig:cmmfaro} illustrates the CMM and FARO Arm machines that we used throughout this study for the experimentation.

Modern CMM technology represents a foundational element of precision measurement in manufacturing. CMMs continue to show advancements in precision measurement, while FARO Arms are more popular to researchers due to their portability. Some studies suggest that if the environmental sensitivity issues are solved in FARO Arms, then it might boost accuracy to the level of CMMs~\cite{ali2016experimental}.

DT's application of real-time monitoring and precision aligns with metrology, which draws attention to several researchers. For instance, Lindqvist et al. (2023) introduced simulation-based optimization algorithms for large-volume dimensional metrology, targeting Saab's automated airframe assembly for the eT7-A aircraft. This approach leverages Saab’s internal case study data and DT simulations to increase measurement accuracy and efficiency~\cite{lindqvist20233d}. Feng et al. (2023) applied DTs in offshore oil production to develop efficient maintenance schedules, lower production costs, and keep production stable~\cite{feng2023multi}.

Adding Internet of Things (IoT) ecosystems has improved DTs' capabilities. IoT allows continuous monitoring by sending data from sensors and machines to DTs. Liu et al. (2023) demonstrated how IoT supports predictive maintenance in manufacturing by analyzing equipment performance and potential failures in real time. Throughout the study, the authors were able to reduce disruptions and boost production efficiency~\cite{liu2023evaluative}.

DTs can be used to simulate and track production lines in the automotive and similar industries, as they help detect possible errors before becoming a major problem, making necessary system corrections. This ensures a high-quality product while maintaining correct system operation. Cavalcanti et al. (2023) also recommended that DT can improve fault prediction and performance of the system using a virtual environment with real-time monitoring~\cite{cavalcanti2023digital}.

However, real-time data is tough with DTs. It needs CPU adversarial power to align and handle different data types for communication in between of a system. Hildebrandt et al. (2023) recommended that the integration of external data sources (e.g., cameras or LiDAR), which could be automated, would provide for an improved optimization but lead to higher system complexity at the same time~\cite{hildebrandt2023automated}.

Machine Learning (ML) methods have recently become quite popular in both industry and academia. For instance, Kerkeni et al. (2024) demonstrated that ML could be used to forecast system functioning, which in turn leads to less downtime and power savings [232]. In another study by Perno et al. (2023), it was also proven that using ML techniques in DT can provide better prediction of operational parameters compared to traditional methods for processes modeled through catalyst manufacturing. Manufacturers can use ML models that were trained on massive data sets to maximize production schedules, reduce waste, and gain the upper hand in manufacturing~\cite{perno2023machine}.

Hybrid metrology systems are beginning to appear, combining CMM and portable technologies, such as 3D scanning and Computed Tomography (CT), into one system that further helps manufacturers with flexibility and precision in recent years. Such systems can combine the best of various technologies; CMMs provide a high level of precision, whereas 3D scanning and CT offer portability and real-time feedback. Considering the importance of surface and dimensional quality in additive manufacturing, hybrid systems are recommended for achieving a desirable output that is suitable to use based on design specifications. A study by Loyda et al. (2023) demonstrated the capabilities of hybrid systems for measuring distortions and roughness in complex shapes; they combined different measurement technologies, providing an overall quality monitoring solution~\cite{loyda2023meeting}.

While several studies explore diverse approaches in metrology, each has limitations, and none of the studies use a combined approach where metrology, DT, and ML are incorporated to develop digital manufacturing models. Apart from this, studies that utilize ML solutions to detect manufacturing defects lack integration with statistical methods for comprehensive uncertainty quantification. Therefore, combining advanced metrology systems with DT frameworks, supported by ML and real-time analytics, will reshape manufacturing processes. Thus, there is a need to develop advanced metrology in manufacturing, which can improve measurement and predictive accuracy with fewer resources and excellent operational efficiency. Integrating DT, metrology, and ML could bridge these gaps, allowing real-time data processing, predictive accuracy, and adaptive environmental control. 

\subsection{Objectives of the Study}

\begin{figure}[h]
    \centering
    \includegraphics[width=\linewidth]{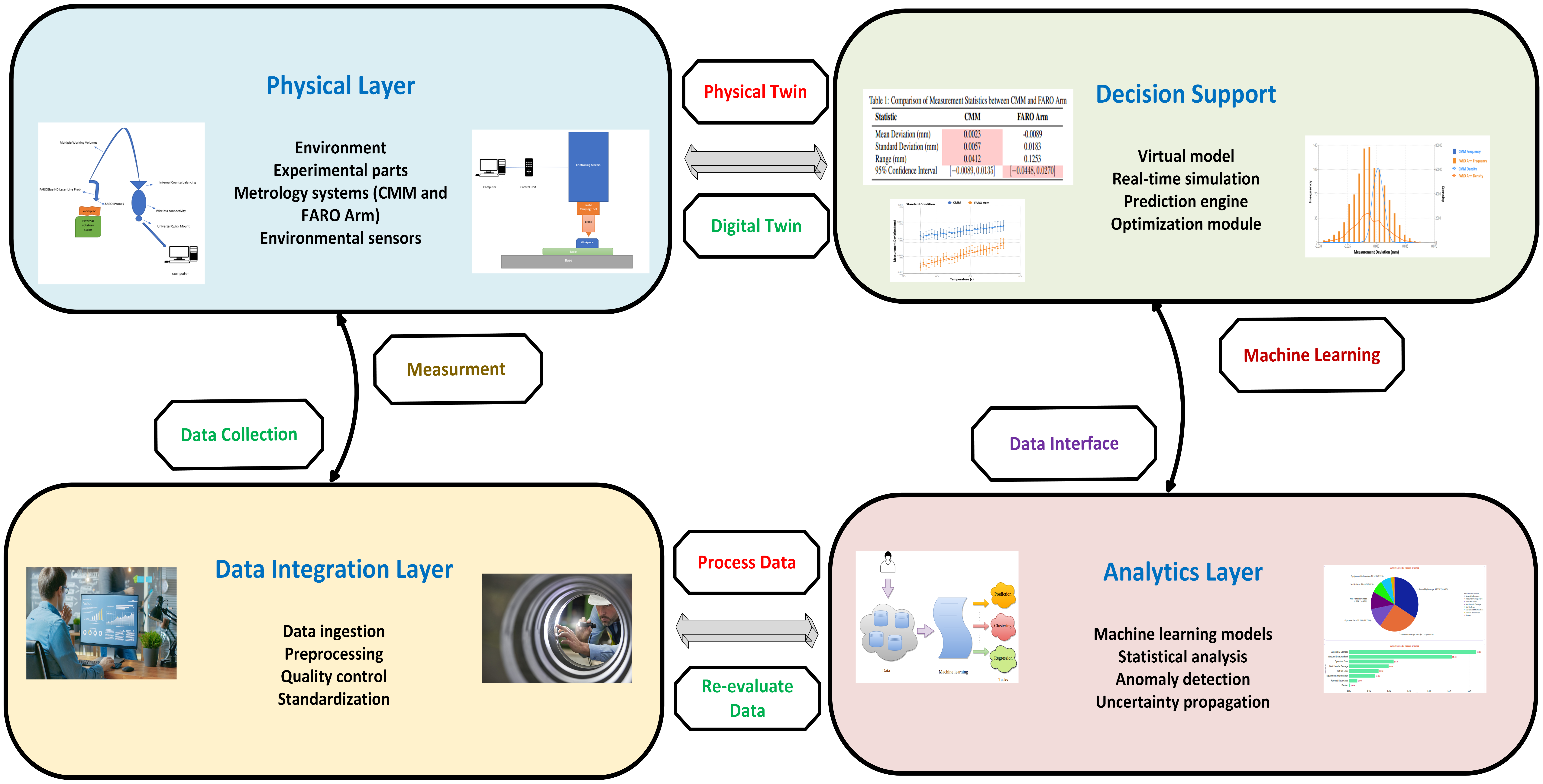}
    \caption{Proposed Metrology and Manufacturing-Integrated Digital Twin (MM-DT) framework.}
    \label{fig:metrodt}
\end{figure}

Considering the opportunity to bridge the gap between DT technology and metrology, In this work, we have proposed a novel DT framework, namely Metrology and Manufacturing-Integrated Digital Twin (MM-DT). As illustrated in Figure~\ref{fig:metrodt}, the proposed MM-DT framework integrated physical data into a DT, and ML-based approaches are then used to analyze the predictive models and develop decision support systems.

Our technical contributions are summarized below:
\begin{itemize}
    \item Analyze and evaluate the measurement characteristics of CMM and FARO Arm devices across various parts and environmental conditions.
    \item Propose the MM-DT framework for integrating metrology data into DT models, enabling real-time quality control and predictive maintenance.
    \item Investigate the impact of environmental factors on measurement accuracy and precision, particularly temperature.
    \item Implement ML-based approaches to improve the predictive performance of DT models on metrology data.
\end{itemize}

The remainder of the paper is organized as follows: the methodology of the study is described in Section~\ref{methods}, the results are presented in Section~\ref{observation}, a brief analysis of the findings is detailed in Section~\ref{discussion}, and the overall contributions of the research are summarized in Section~\ref{conclusions}.
\section{Methodology}\label{methods}
\subsection{Experimental Setup}
We have designed a comprehensive experimental setup to develop the MM-DT framework that includes various manufacturing components. As an effect, Twenty parts with varying geometries and sizes were chosen to represent, which includes simple shapes such as cylinders, cubes, spheres, and complex geometries like turbine blades and gear assemblies. 

We have used two types of measurement devices, CMM and FARO Arm. The CMM is a high-precision device with an accuracy level of $\pm (1.5 + \frac{L}{333}) \, \mu m$, where "L" is the length measured in millimeters. It is a touch-sensitive probe and needs to operate at a controlled room temperature~\cite{hocken2012coordinate}. On the other hand, the FARO Quantum S FaroArm is a portable measuring device with an accuracy of $\pm 0.036$ millimeters~\cite{chanda2014design}.

To examine the impact of temperature variations on measurement accuracy, all measurements were conducted in a controlled environment. Two temperature settings were used: a) Standard condition: $20^\circ \mathrm{C} \pm 0.5^\circ \mathrm{C}$ ($68^\circ \mathrm{F} \pm 0.9^\circ \mathrm{F}$) and b) Elevated condition: $30^\circ \mathrm{C} \pm 0.5^\circ \mathrm{C}$ ($86^\circ \mathrm{F} \pm 0.9^\circ \mathrm{F}$). Humidity was maintained at $50\% \pm 5\%$ for all measurements to minimize its impact on the results.

Each part was measured 2 times on the CMM and FARO Arm set at each of these temperatures, so eight measurements
 per configuration were taken during this study. Key dimensions such as length, width, diameter, and angle were
 recorded depending on the geometry of each part. At the other end of complexity, we added flatness and cylindricity
 measurements. Experimental bias was minimized by randomizing the measurement sequence. In addition to this, a standardized fixturing system was used for uniform part positioning throughout all measurements. Moreover, every part was conditioned at the target temperature for more than 4 hours before each measurement to attain its stable thermal
 state. In order to prepare the CMM and FARO Arm for accurate, reliable measurements, both were aligned using
 certified reference standards from their respective national metrology institutes’ guidelines. Before each data collection
 session, calibration checks were conducted to ensure the measurement systems were stable.

\subsection*{Data Collection and Analysis Approaches}
We collected the following data for each measurement: Part-Identifier \& -Description, Measurement Device (CMM or FARO Arm), Temperature Condition, Nominal Value of Dimension, Measured Value of Dimension and Deviation from Nominal-Value to Specified Tolerance Band., Date-Time when Measured, Operator-ID and Environmental Data (Temperature / Humidity) aside apprometric Measure Duration. 320 measurements were recorded (20 parts × 2 devices × 2 temperatures × 2 repetitions). 

\begin{figure}
    \centering
    \includegraphics[width=\linewidth]{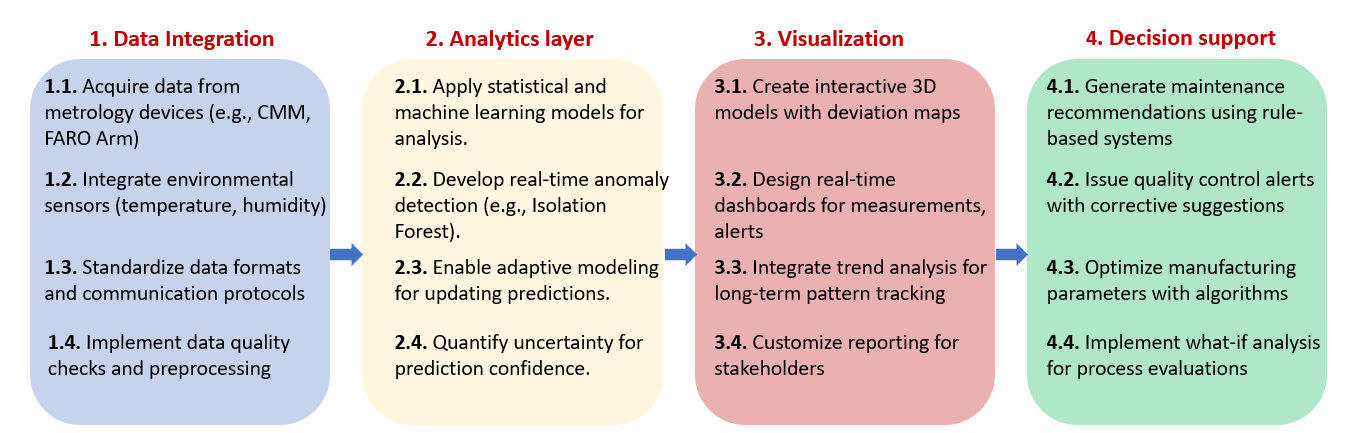}
    \caption{Framework for integrating metrology data into a DT system, consisting of four key stages: (1) Data Integration, where data is acquired from devices (e.g., CMM, FARO Arm) and environmental sensors, followed by standardization and preprocessing; (2) Analytics Layer, which applies statistical and machine learning models, enables real-time anomaly detection (e.g., Isolation Forest), adaptive modeling, and uncertainty quantification; (3) Visualization, where interactive 3D models, real-time dashboards, and trend analysis tools are used to track and report data; and (4) Decision Support, which provides maintenance recommendations, quality control alerts, parameter optimization, and "what-if" process analysis.}
    \label{fig:dtingestframe}
\end{figure}

After analyzing the collected data, several statistical and ML models were proposed. Significance testing was performed using statistical methods, including mean and standard deviations analysis, paired t-tests, and ANOVA to examine the significant differences between CMM and FARO Arms at different temperatures. ML models like Random Forest, Support Vector Regression, Gradient Boosting, and Neural Networks are used to predict measurement deviations on the specific machine. In contrast, Isolation Forest is used for anomaly detection. We have used five-fold cross-validation in all our experiments to train and test the ML models with evaluation metrics such as $R^2$, MAE, RMSE, and F1-score.
Finally, we ingested analyzed metrology data into DT modes, as shown in Figure~\ref{fig:dtingestframe}.
\section{Results and Analysis}\label{observation}
\subsection{Measurement Variability Analysis}
Our analysis of the measurement data revealed significant insights into the performance characteristics of both the CMM and FARO Arm devices across various parts and environmental conditions. Table~\ref{tab:tab1} summarizes the key statistics for CMM and FARO devices. From the table, it can be observed that the CMM consistently outperformed the FARO Arm device in terms of precision and accuracy. The CMM exhibited a smaller mean deviation from the nominal values (0.0023 mm) compared to the FARO Arm (-0.0089 mm). Moreover, the standard deviation of CMM measurements (0.0057 mm) was significantly lower than that of the FARO Arm (0.0183 mm), indicating higher precision.

\begin{table}[h!]
    \centering
    \caption{Comparison of Measurement Statistics between CMM and FARO Arm.}
    \begin{tabular}{l>{\centering}p{3cm}>{\centering\arraybackslash}p{3cm}}
        \toprule
        \textbf{Statistic} & \textbf{CMM} & \textbf{FARO Arm} \\
        \midrule
        Mean Deviation (mm) & \cellcolor{red!20}0.0023 & -0.0089 \\
        Standard Deviation (mm) & \cellcolor{red!20}0.0057 & 0.0183 \\
        Range (mm) & \cellcolor{red!20}0.0412 & 0.1253 \\
        95\% Confidence Interval & \([-0.0089, 0.0135]\) & \cellcolor{red!20}\([-0.0448, 0.0270]\) \\
        \bottomrule
    \end{tabular}
    \vspace{0.3cm}
    \caption*{\textit{Note:} Red shading indicates the best-performing measurement in each category. Lower values for mean deviation, standard deviation, and range indicate better performance, while confidence intervals closer to zero suggest greater measurement reliability.}
    \label{tab:tab1}
\end{table}

\begin{figure}[h!]
    \centering
    \includegraphics[width=\linewidth]{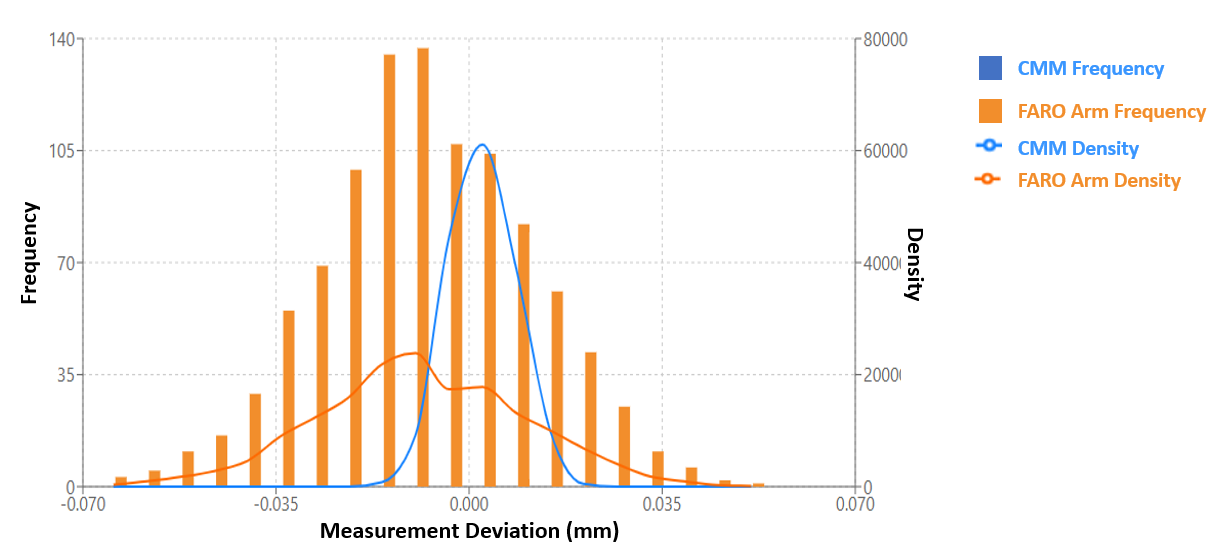}
    \caption{Comparison of measurement deviations between CMM and FARO Arm.}
    \label{fig:figcmm}
\end{figure}

The improved visualization effectively highlights the performance differences between the two measurement devices, providing a statistically robust representation of the data. Figure~\ref{fig:figcmm} shows the comparison between CMM (blue) and FARO Arm (orange), indicating the precision and accuracy characteristics of both devices. The CMM displays a narrower, more peaked distribution centered close to zero, suggesting higher precision and marginally better accuracy. In contrast, the FARO Arm's wider and flatter distribution points to relatively lower precision, but still acceptable accuracy for many applications. The broader range of deviations in the FARO Arm could be attributed to its portable nature.
\subsection{Impact of Environmental Factors}

To assess the influence of temperature on measurement accuracy, we conducted a multiple linear regression analysis. Table~\ref{tab:regcoeff} presents the model coefficients, indicating that temperature and device type significantly impact measurement deviation. The p-values, such as 0.0023 for the intercept and 0.0015 for the temperature coefficient, suggest that these effects are statistically significant. 
\begin{table}[h!]
    \centering
    \caption{Regression model coefficients.}
    \label{tab:regcoeff}
    \begin{tabular}{@{}lcc@{}}
        \toprule
        \textbf{Variable} & \textbf{Coefficient} & \textbf{p-value} \\ 
        \midrule
        Intercept         & -0.0152              & 0.0023           \\
        Nominal           & 0.00015              & $<$ 0.001        \\
        Device            & 0.0112               & $<$ 0.001        \\
        Temperature       & 0.00078              & 0.0015           \\
        \bottomrule
    \end{tabular}
\end{table}

\begin{figure}
    \centering
    \includegraphics[width=.8\linewidth]{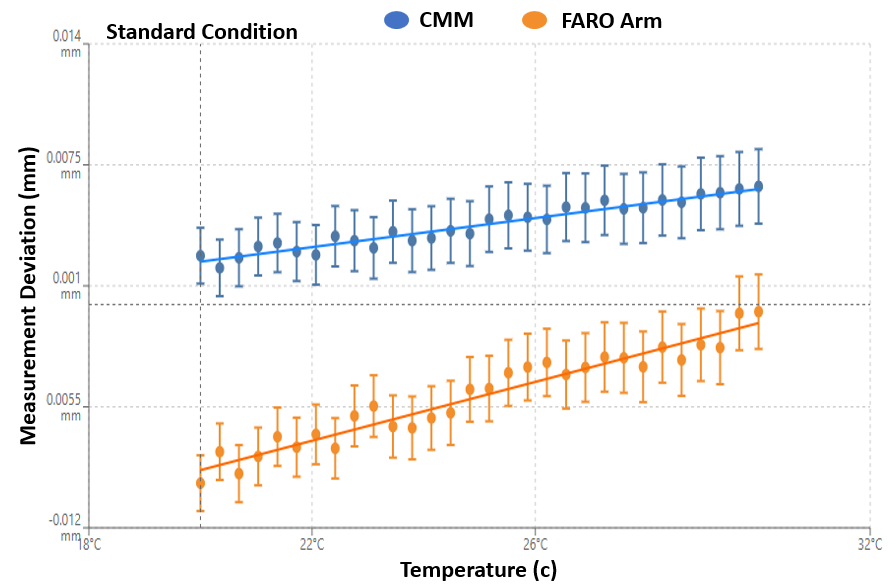}
    \caption{Impact of temperature on measurement deviations for CMM and FARO Arm.}
    \label{fig:figtemp}
\end{figure}

Figure~\ref{fig:figtemp} visualizes these relationships, showing how deviations systematically vary with temperature for both the CMM and FARO Arm. This combined analysis emphasizes the importance of accounting for temperature and device-specific characteristics when evaluating and ensuring measurement accuracy.
\subsection{Machine Learning Model Performance}
We implemented several machine learning models to predict measurement deviations based on part characteristics, device type, and environmental conditions. 
Table~\ref{tab:performance_metrics} provides a comparative analysis of five ML models' performance considering five different metrics used during this study. From the table, it can be observed that the Ensemble model (RF + GB) achieved the best performance, with the highest R\textsuperscript{2} score of 0.91 and the lowest RMSE and MAE values (1.59 $\mu$m and 1.15 $\mu$m), highlighted in bold red. Conversely, the Support Vector Machine model showed the least performance with the lowest R\textsuperscript{2} score of 0.85 and RMSE (1.89 $\mu$m) and MAE (1.45 $\mu$m), highlighted in bold blue. 

\begin{table}[h!]
    \centering
    \caption{Comparative analysis of individual model performance metrics.}
    \label{tab:performance_metrics}
    \begin{tabular}{@{}lccc@{}}
        \toprule
        \textbf{Model Type} & \textbf{R\textsuperscript{2} Score} & \textbf{RMSE ($\mu$m)} & \textbf{MAE ($\mu$m)} \\
        \midrule
        Random Forest (RF) & 0.89 & 1.68 & 1.23 \\
        Gradient Boosting (GB) & 0.88 & 1.74 & 1.31 \\
        \textbf{\textcolor{blue}{Support Vector Regression}} & \textbf{\textcolor{blue}{0.85}} & \textbf{\textcolor{blue}{1.89}} & \textbf{\textcolor{blue}{1.45}} \\
        Neural Network (MLP) & 0.86 & 1.82 & 1.38 \\
        \textbf{\textcolor{red}{Ensemble (RF + GB)}} & \textbf{\textcolor{red}{0.91}} & \textbf{\textcolor{red}{1.59}} & \textbf{\textcolor{red}{1.15}} \\
        \bottomrule
    \end{tabular}
\end{table}

\subsection{Anomaly Detection}
We implemented an Isolation Forest algorithm to detect anomalies in measurement deviations, aiming to identify unusual measurements that could indicate part defects or errors. Table~\ref{tab:anomaly} summarizes the performance metrics of the anomaly detection system. The system achieved a high true positive rate of 0.92 and a low false positive rate of 0.03, resulting in an F1 score of 0.89. 
\begin{table}[h!]
    \centering
    \caption{Isolation Forest anomaly detection performance.}
    \label{tab:anomaly}
    \begin{tabular}{@{}lc@{}}
        \toprule
        \textbf{Metric} & \textbf{Value} \\ 
        \midrule
        Contamination Factor & 0.05 \\
        False Positive Rate & 0.03 \\
        True Positive Rate & 0.92 \\
        F1 Score & 0.89 \\
        \bottomrule
    \end{tabular}
\end{table}

\begin{figure}[h]
    \centering
    \includegraphics[width=\linewidth]{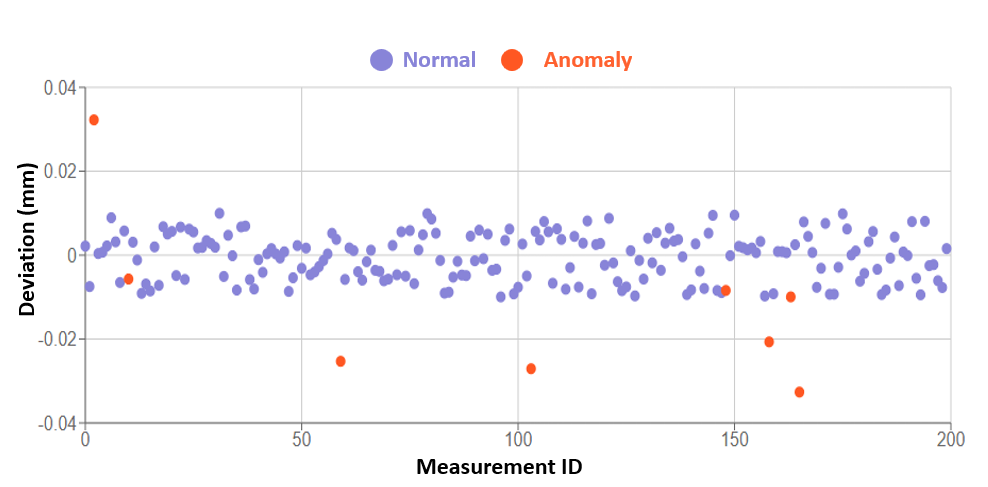}
    \caption{Anomaly detection in measurement deviations using Isolation Forest.}
    \label{fig:isoram}
\end{figure}

Figure~\ref{fig:isoram} visually differentiates detected anomalies from normal measurements, enabling rapid identification of potential quality issues. This visualization aids in continuous process improvement and robust digital twin modeling.
\subsection{Digital Twin Framework Performance}
To evaluate the effectiveness of our proposed digital twin framework, we implemented a continuous learning pipeline that periodically retrains the predictive models with new measurement data. Table~\ref{tab:retrain} shows the impact of different retraining intervals on model performance. The results indicate that more frequent retraining leads to consistent improvements, with the quarterly retraining schedule showing the largest gains in average R\textsuperscript{2} improvement (0.09) and RMSE reduction (0.52~$\mu$m).

\begin{table}[h!]
    \centering
    \caption{Model retraining schedule and performance improvement.}
    \label{tab:retrain}
    \begin{tabular}{@{}lcc@{}}
        \toprule
        \textbf{Retraining Interval} & \textbf{Avg. R\textsuperscript{2} Improvement} & \textbf{Avg. RMSE Reduction ($\mu$m)} \\ 
        \midrule
        Weekly & 0.02 & 0.15 \\
        Monthly & 0.05 & 0.31 \\
        Quarterly & \textbf{\textcolor{red}{0.09}} & \textbf{\textcolor{red}{0.52}} \\
        \bottomrule
    \end{tabular}
\end{table}

\begin{figure}[h]
    \centering
    \includegraphics[width=\linewidth]{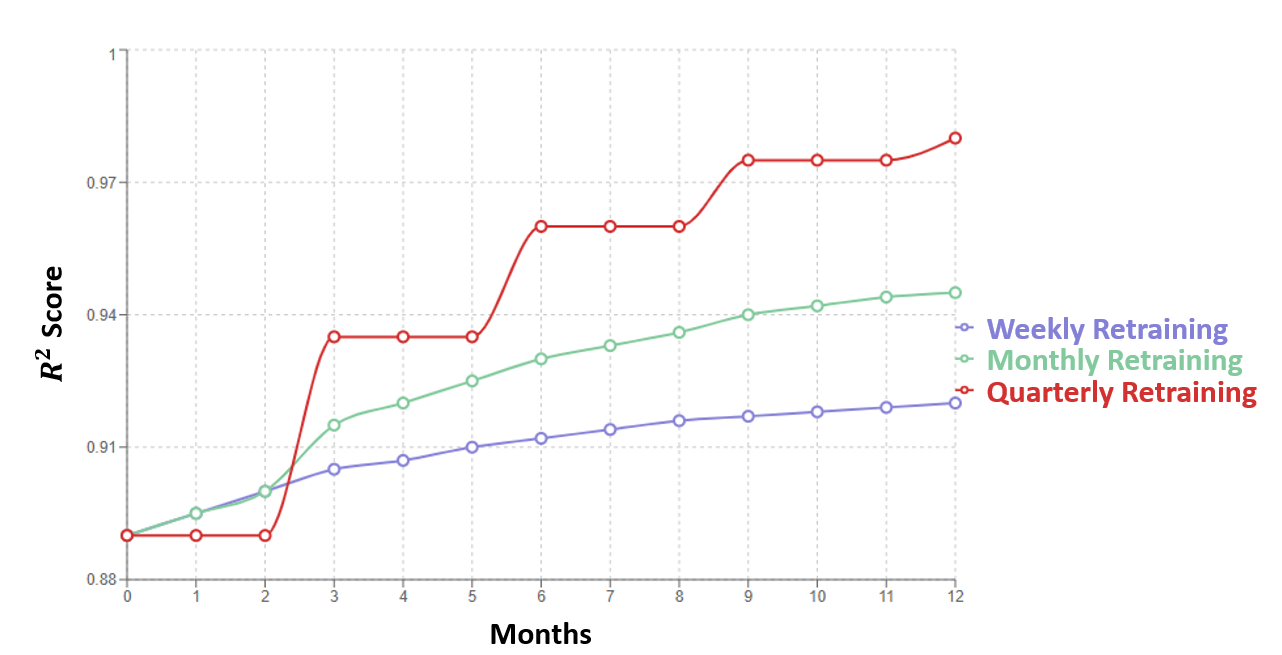}
    \caption{Comparison of different retraining schedules.}
    \label{fig:difre}
\end{figure}

Figure~\ref{fig:difre} effectively highlights these nuances over a full year with clear x-axis labeling. It shows the stepped nature of improvements for less frequent retraining and more gradual improvement for frequent retraining. This balance between up-to-date models and computational efficiency is essential for optimizing continuous learning pipelines in digital twin models.

\subsection{Continuous Learning Pipeline Performance}
\begin{figure}[h!]
    \centering
    \includegraphics[width=0.8\textwidth]{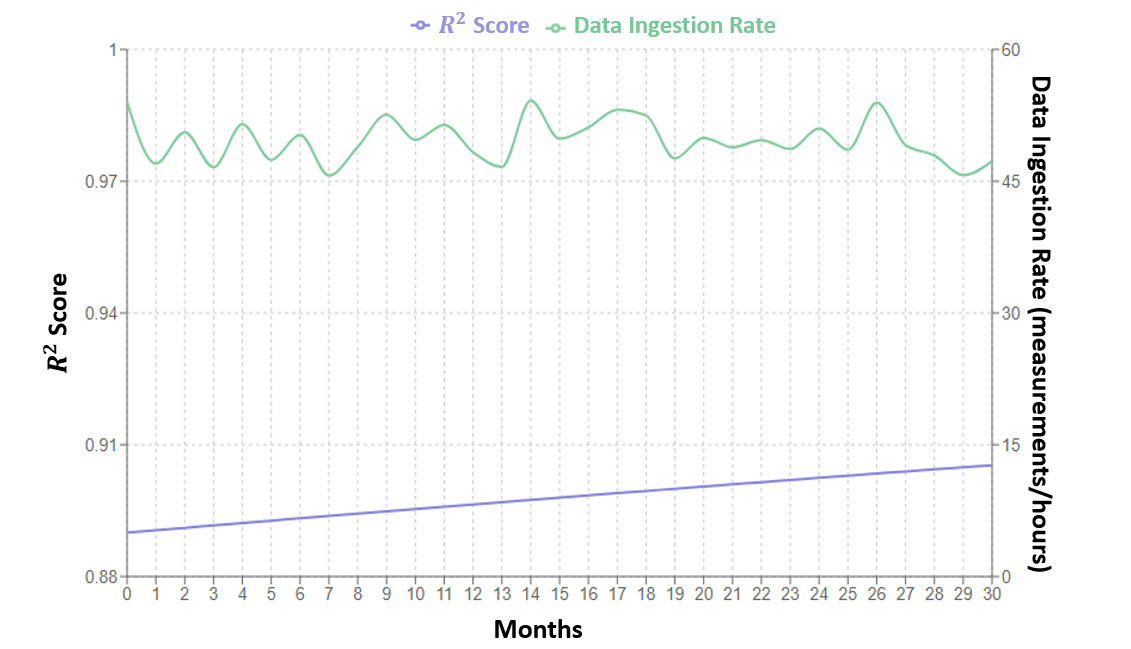} 
    \caption{Continuous learning pipeline performance in DT framework.}
    \label{fig:fig7}
\end{figure}

Figure~\ref{fig:fig7} highlights the effectiveness of the continuous learning pipeline within our DT framework. The steady improvement in the R\textsuperscript{2} score, as shown in the figure, demonstrates the system’s ability to continuously adapt and refine itself, ensuring the DT remains accurate over time. As shown in Table~\ref{tab:pipeline} the system can ingest 50 measurements per hour and update the model every 24 hours, with each update contributing an average improvement of 0.005 in the R\textsuperscript{2} score.

\begin{table}[h!]
    \centering
    \caption{Continuous learning pipeline performance metrics.}
    \label{tab:pipeline}
    \begin{tabular}{@{}lc@{}}
        \toprule
        \textbf{Metric} & \textbf{Value} \\ 
        \midrule
        Data Ingestion Rate (measurements/hour) & 50 \\
        Model Update Frequency/hours & 24 \\
        Average Model Convergence Time/minutes & 45 \\
        Improvement in R\textsuperscript{2} Score & 0.005 \\
        \bottomrule
    \end{tabular}
\end{table}
\section{Discussion}\label{discussion}
Our study highlights the importance of seamless data integration for accurate DT modeling. By combining data from CMM and FARO Arm measurements within a single framework, we establish a cohesive data source that supports comprehensive analysis and prediction. This integration allows MM-DT to maintain a holistic view of part measurements, especially when devices are used under different conditions. 

In our experiment, we found that the CMM provides high precision in measurements, with a mean deviation of 0.0023 mm and a standard deviation of 0.0057 mm, compared to the FARO Arm (-0.0089 mm mean deviation, 0.0183 mm standard deviation). This indicates that the CMM is better suited for tasks that require high accuracy.

We also found that the FARO Arm is sensitive to temperature changes. Our analysis showed that temperature significantly affected FARO Arm measurements. While the FARO Arm provides flexibility for field applications, we found that it might be necessary to adjust the temperature occasionally.

In this study, we have used ML models in the MM-DT framework to improve prediction accuracy. Our results indicated that an Ensemble model (RF + GB) surpassed previous ML models, attaining a $R^2$ score of 0.91, an RMSE of 1.59 µm, and an MAE of 1.15 µm in forecasting measurement errors. This predictive capability helps enhance decision-making and quality control in real-time. Furthermore, we employed the Isolation Forest method for anomaly detection, with an accuracy of 0.92 and an F1 score of 0.89. A continuous learning process has been incorporated into MM-DT to ensure models remain current. This pipeline refreshes model parameters every 24 hours. Quarterly retraining of the model enhances accuracy, with an average $R^2$ gain of 0.09 and a reduction in RMSE of 0.52 µm. This real-time adaptation enables MM-DT to continually respond to fresh data. Thus, our continuous learning pipeline is an essential approach to maintaining accuracy in dynamic manufacturing settings.
\subsection{Limitations and Future Work}

Our study identified several limitations within the MM-DT framework and areas for potential future work to address these gaps. First, we realized that measurement uncertainty remains a critical factor in predictive accuracy. Variations in predictive performance might result from minor inconsistencies in device calibration. Addressing this issue through better calibration approaches or uncertainty quantification models may improve the reliability of MM-DT. Therefore, in our future work, we will focus on minimizing uncertainty to ensure consistent predictions, even under fluctuating conditions.

Second, our anomaly detection efforts with the Isolation Forest algorithm show promising results. However, we primarily applied anomaly detection to simpler geometries and measurement types. Expanding this by including complex parts with intricate geometrical features might increase the robustness of the MM-DT framework. Therefore, future studies should focus on more advanced anomaly detection techniques by incorporating deep learning-based approaches with different manufacturing components.

Another limitation lies in the computational demands of real-time model updates. The continuous learning pipeline in MM-DT, which updates model parameters every 24 hours, requires considerable computational resources, especially when processing large amounts of data. To prevent potential lags, it is necessary to balance the update frequency with computational efficiency. In future work, we will consider using advanced GPU systems and cloud-based storage systems to optimize the operational procedure.

Finally, MM-DT’s current device integration is limited to CMM and FARO Arm data. Many manufacturing environments use a variety of metrology tools, such as laser scanners, optical systems, and non-contact measurement technologies, which could complement CMM and FARO Arm data. Thus, broadening MM-DT to include these tools would make it a more advanced and adaptable DT framework in manufacturing domains.

\section{Conclusion}\label{conclusions}
In this study, we introduced the MM-DT framework, which measures, analyzes metrology data from CMM and FARO Arms, and integrates with DT technology. Our findings suggest that CMM offers better precision measurement, while the FARO Arm provides flexibility but is more sensitive to environmental changes. By integrating Ensemble ML approaches, combining Random Forest and Gradient Boosting, we were able to significantly improve the predictive performance of MM-DT, achieving an $R^2$ score of 0.91. The use of a continuous learning pipeline further ensures MM-DT’s adaptability to real-time data processing. Throughout this process, we showed unique approaches where DT, metrology, and ML-based approaches are combined and used for analysis of manufacturing data. However, measurement uncertainty remains an ongoing issue, which will be addressed in our future work. Additionally, future work will focus on improved computational efficiency with high GPU resources and genetic algorithm approaches, physics-guided ML-based approaches to develop more generalized and robust data-driven DT techniques, and also the use of various metrology tools tested on various manufacturing settings to develop a more standard DT framework.

\section*{Conflict of interest}
The authors declare no conflict of interest.
\bibliographystyle{unsrt}  
\bibliography{main}

\end{document}